\documentclass[journal]{IEEEtran}
\usepackage{graphicx}
\usepackage{a4}
\usepackage{fullpage}
\usepackage{epsfig}
\usepackage{epsf}
\usepackage{graphics}
\usepackage{amsmath}
\usepackage{tabls}
\usepackage{amssymb}
\usepackage{verbatim}
\usepackage[latin1]{inputenc}
\usepackage{subfigure} 
\usepackage{fancyhdr}
\usepackage{cite}
\usepackage{ulem}
\usepackage{subfigure}
\usepackage{lipsum}
\usepackage{algorithm}
\usepackage{algorithmic}
\usepackage[space]{grffile}
\usepackage{array}
\usepackage{xcolor}
\usepackage{mathtools}
\usepackage{multirow}

\begin{document}
\title{Experimental characterization of Raman amplifier optimization through inverse system design}

\author{Uiara~C.~de~Moura,
        Francesco~Da~Ros,
        A.~Margareth~Rosa~Brusin,
        Andrea~Carena, 
        and~Darko~Zibar
\thanks{U.~C.~de~Moura, F.~Da~Ros and D.~Zibar are with DTU~Fotonik, Department of Photonics Engineering, Technical University of Denmark, Kgs. Lyngby, Denmark. (e-mail: uiamo@fotonik.dtu.dk).}
\thanks{A.~M.~Rosa~Brusin and A.~Carena are with Dipartimento di Elettronica e Telecomunicazioni (DET), Politecnico di Torino, Torino, Italy}
\thanks{Manuscript received xx xx, 20xx; revised xx xx, 20xx.}}



\maketitle

\begin{abstract}
Optical communication systems are always evolving to support the need for ever--increasing transmission rates. This demand is supported by the growth in complexity of communication systems which are moving towards ultra--wideband transmission and space--division multiplexing. Both directions will challenge the design, modeling, and optimization of devices, subsystems, and full systems. Amplification is a key functionality to support this growth and in this context, we recently demonstrated a versatile machine learning framework for designing and modeling Raman amplifiers with arbitrary gains. In this paper, we perform a thorough experimental characterization of such machine learning framework. The applicability of the proposed approach, as well as its ability to accurately provide flat and tilted gain--profiles, are tested on several practical fiber types, showing errors below 0.5~dB. Moreover, as channel power optimization is heavily employed to further enhance the transmission rate, the tolerance of the framework to variations in the input signal spectral profile is investigated. Results show that the inverse design can provide highly accurate gain--profile adjustments for different input signal power profiles even not considering this information during the training phase.

\end{abstract}

\begin{IEEEkeywords}
optical communications, optical amplifiers, machine learning, neural networks.
\end{IEEEkeywords}

\IEEEpeerreviewmaketitle

\section{Introduction}
\label{sec:introduction}
\IEEEPARstart{O}{ptical} amplifiers are key devices in optical communication systems, with the erbium--doped fiber amplifier (EDFA) as the most deployed technology in commercial systems\cite{agrell2016roadmap}. EDFAs were responsible for the multi--channel transmission revolution on the 90's and now are one of the main bottlenecks to realize transmission systems beyond C and L bands. Raman amplifiers have been investigated as an alternative to realize such systems since they are naturally broadband and able to provide gain at any wavelength~\cite{Iqbal19}. Most importantly, they are flexible to shape the gain--profile by properly adjusting the pump power and wavelengths when operating in multi--pump configuration. This is a critical property for future ultra--wideband systems, since the channel power profiles that maximize the achievable information rate (AIR) are not necessarily flat due to increased levels of Kerr nonlinearity and stimulated Raman scattering~\cite{Hamaoka19,Ferrari19}.

Therefore, a problem that have gained renewed interest is the Raman amplifier inverse design. It consists in finding the laser pump configuration (power and wavelength) for a desired Raman gain spectral profile. Conventionally, the Raman amplifier inverse design requires solving a set of nonlinear ordinary differential equations (ODEs) that govern the complex pump--signal, signal--signal and pump--pump interactions during their propagation in the optical fiber. Therefore, it is a time--consuming and complex optimization process, especially for counter--propagating pump scheme, which is a better approach in terms of noise figure \cite{agrawalRaman}. Over the years, this problem has been addressed by global optimization algorithms such as evolutionary algorithms~\cite{Ferreira11, Zhou01, Chen18, Perlin02} and particle swarm optimization~\cite{Mowla08, Jiang10}. They are fastened by analytical~\cite{Ferreira11, Zhou01}, numerical~\cite{Mowla08, Jiang10} or even artificial neural network~\cite{Zhou06,Chen18} models that approximate the Raman amplifier's ODEs solution. Some proposals break the design problem into two simpler inverse problems: firstly finding the pump wavelengths using genetic algorithm and secondly finding the pump powers iteratively solving the ODEs~\cite{Perlin02}. Others reduce the parameters to be adjusted aiming at simplifying the inverse design problem by adjusting groups of laser pumps instead of each one individually~\cite{Emori01, Ania07}.

All these approaches completely rely on optimization loops that, even fastened by approximations and simplifications, require several iterations to provide the pump configuration. Moreover, such optimizations need to be restarted for every new target gain--profile. Therefore, the development of new tools to reduce the Raman amplifier inverse design complexity is essential for dynamic optical networks targeting near--real--time adaptation against physical layer changes~\cite{pointurier2019introduction}.

To avoid time--consuming optimization loops, an inverse system design based on machine learning has been recently applied to the Raman amplifier case~\cite{Zibar:19,Zibar20,Ionescu19,deMoura20,Ye20,ourJLTonArXiv}. These works demonstrated that an artificial neural network (NN) can learn the inverse mapping of the Raman amplifiers. This inverse mapping provides the pump configuration (power and wavelength) as a function of the Raman gain spectral profile. Once properly trained, the same inverse mapping NN can be applied for any new target gain--profile, promptly providing the respective pump configuration. This approach has also been extended to few--mode Raman amplifiers, to simultaneously flatten the gain--profiles and reduce the mode--dependent gain~\cite{Chen:20}.

In this paper, we extend our recent work~\cite{deMoura20}, where the machine learning (ML) framework proposed by~\cite{Zibar:19,Zibar20} is experimentally evaluated in many practical scenarios. In~\cite{deMoura20}, the ML framework was extensively investigated over distributed and discrete counter--propagating Raman amplifiers with different fiber types and lengths. Results show maximum errors between target and designed gain--profiles below 0.5~dB for 80\% of the evaluated cases. As a complementary result for~\cite{deMoura20}, in this work we test the ML framework ability in achieving flat and tilted gain--profiles. Results show a maximum of 0.5~dB of deviation from target gain--profiles for all investigated Raman amplifiers.

Additionally, we evaluate the impact of different input power profiles on the NN models that build the ML framework. This is done by training these models in a data--set with constant input signal power spectral density (PSD) profiles and validate them over different input signal PSDs. This analysis is performed over a 100-km standard single mode fiber (SSMF) distributed Raman amplifier. Although we do not expect high gain--profile dependence for different input signal PSD due to the low inter-signal stimulated Raman scattering (SRS) inside the C--band, results show that the inverse mapping NN is quite sensible to the input signal PSD. In fact, the results show some degradation in predicting the pump powers, with a maximum error of 90~mW, when evaluated over different input signal PSDs. Results also show that this degradation can be overcome by either applying the gradient descent (GD)--based fine--optimization routine or considering the information on the input signal PSD as an additional input on the NN models.

The paper is organized as follows. Section~\ref{sec:experimental_setup} presents the experimental setup for the Raman amplifier that provides the data--set to train and validate the NN models of the ML framework. Section~\ref{sec:raman_gain_control} describes the machine learning framework for the inverse system design, detailing the proposed modification to consider input signal PSD information. Section~\ref{sec:results} presents and discussed the experimental results for the ML framework validation when trying to achieve arbitrary, flat and tilted gain--profiles for the different fiber types. Section~\ref{sec:conclusion} concludes this work. 

\section{Experimental setup}
\label{sec:experimental_setup}
\begin{figure*}[!t]
  \centering
  \includegraphics[width=1\textwidth]{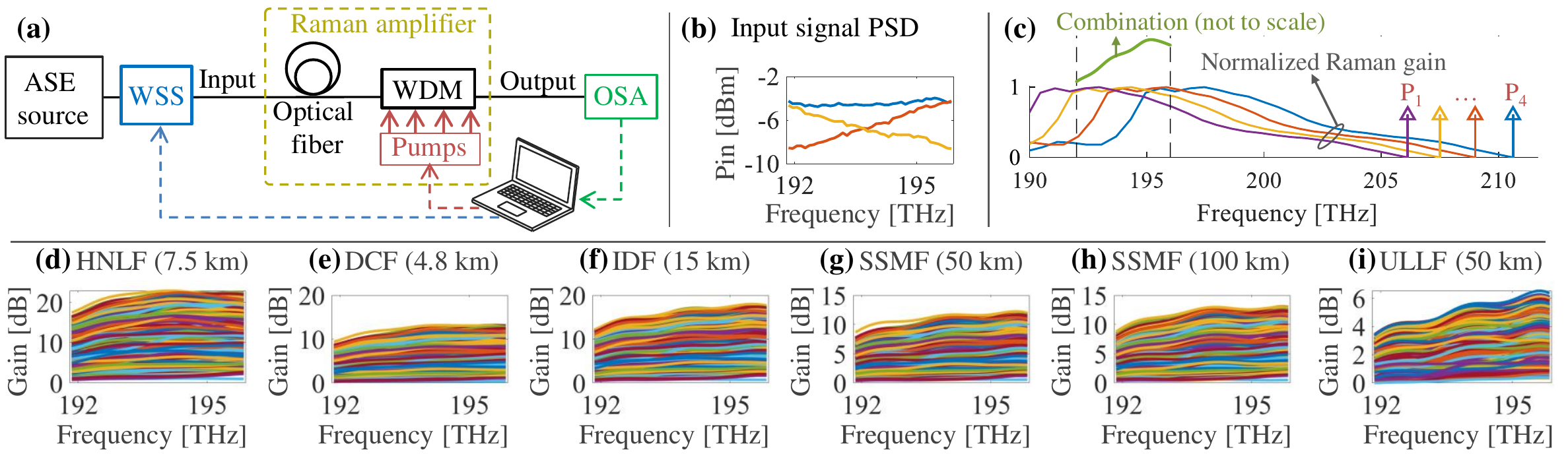}
\caption{(a) Experimental setup to capture the data--set for the NN training and further validations showing the measured on--off gain--profiles for each fiber types/length; (b) some examples of input signal power spectral density (PSD) profiles; (c) normalized Raman gain coefficients for each pump laser; and measured gain--profiles for (d) HNLF, (e) DCF, (f) IDF, (g) SSMF (50~km), (h) SSMF (100~km), and (i) ULLF.}
\label{fig:setup}
\end{figure*}

The experimental setup for the Raman amplifiers is depicted in Fig.~\ref{fig:setup}(a). It is also used for evaluating the performance of the ML framework. An amplified spontaneous emission (ASE) source generates the input signal covering the entire C--band (192-196~THz). A wavelength selective switch (WSS) is used to shape the input signal PSD profile. Some examples of input signal linear PSD profiles are illustrated in Fig.~\ref{fig:setup}(b) for different slopes. The Raman amplifier is composed of an optical fiber and a commercial Raman pump module with four pump lasers. Pump frequencies (shown in Table~\ref{tab:pumps}) are fixed and able to amplify the full C--band. Pump powers are remotely controlled and their maximum values into the optical fiber are also shown in Table~\ref{tab:pumps}. Pumps and signals are combined in a counter--propagating scheme using a wavelength division multiplexer (WDM). At the Raman amplifier output, an optical spectrum analyzer (OSA) measures the signal PSD at a resolution of 0.1~nm. The Raman on--off gain profile is calculated by the difference between the output signal PSDs with the pump lasers turned on and off. This will be the metric used throughout the paper to represent the amplifier gain.

\begin{table}[h]
\centering
\caption{\bf Pump lasers frequency and maximum powers}
\begin{tabular}{ccccc} 
\hline
Pump                & $P_1$ & $P_1$ & $P_1$ & $P_1$ \\
\hline
Frequency (THz)     & 206.1 & 207.5 & 209.0 & 210.6 \\
Maximum power (mW)  & 145   & 158.5 & 180   & 152.5 \\
\hline
\end{tabular}
  \label{tab:pumps}
\end{table}

Five optical fiber types with different characteristics shown in Table \ref{tab:fibers} are considered: 7.5~km of highly nonlinear fiber (HNLF), 4.8~km of dispersion compensating fiber (DCF), 15~km of inverse dispersion fiber (IDF), 50 and 100~km of standard single mode fiber (SSMF), and 50~km of ultra--low loss fiber (ULLF). HNLF, DCF and IDF are special highly nonlinear fibers used for discrete Raman amplifiers~\cite{Iqbal19JLT}. DCF and IDF present the advantage to also compensate for chromatic dispersion. IDF is being used for discrete Raman amplifiers as an alternative to DCFs due to its lower attenuation~\cite{Iqbal19JLT}. SSMF is widely deployed in terrestrial systems while ULLF is used for submarine and unrepeatered links due to its lower loss and wider effective area. These are transmission fibers and they are used to demonstrate the distributed Raman amplification. Total signal power does not exceed 9~dBm at the optical fiber input for the SSMF and ULLF cases and 3~dBm for the HNLF, DCF, IDF cases.

\begin{table}[h]
\centering
\caption{\bf Optical fibers parameters}
\begin{tabular}{cccccc} 
\hline  
Fiber                       & HNLF  & DCF   & IDF   & SSMF  & ULLF \\
\hline
$\alpha_{1550}$ ($dB/km$)   & 1.0  & 0.5  & 0.23 & 0.2  & 0.16  \\
$\alpha_{1450}$ ($dB/km$)   & 1.2  & 0.8  & 0.31 & 0.25 & 0.2   \\
$A_{eff}$ ($\mu m^2$)       & 10   & 15   & 31   & 80   & 153   \\
$g_R$ ($W^{-1} km^{-1}$)    & 6.3  & 3    & 1.3  & 0.8  & 0.52  \\
\hline
\end{tabular}
\vspace{1ex}
{\\ \raggedright $\alpha$: attenuation, $A_{eff}$: effective area, $g_R$: Raman gain coefficient. \par}
  \label{tab:fibers}
\end{table}

Fig.~\ref{fig:setup}(c) shows approximations for the normalized Raman gain coefficients for each pump laser. More accurate curves should be scaled according to~\cite{Rottwitt:s}. Fig.~\ref{fig:setup}(d-i) shows the measured gain--profiles for each fiber type and different pump powers. High frequency signals in Fig.~\ref{fig:setup}(d-i) have higher gains due to the additive contribution of each pump laser, as illustrated in Fig.~\ref{fig:setup}(c). This is because on the C--band, channels are close enough ($<$~4THz) and therefore they are not strongly affected by the stimulated Raman scattering (SRS) power transfer from high to low frequency channels.

\section{Raman gain--profile control}
\label{sec:raman_gain_control}
\begin{figure*}[!t]
  \centering
  \includegraphics[width=1.0\textwidth]{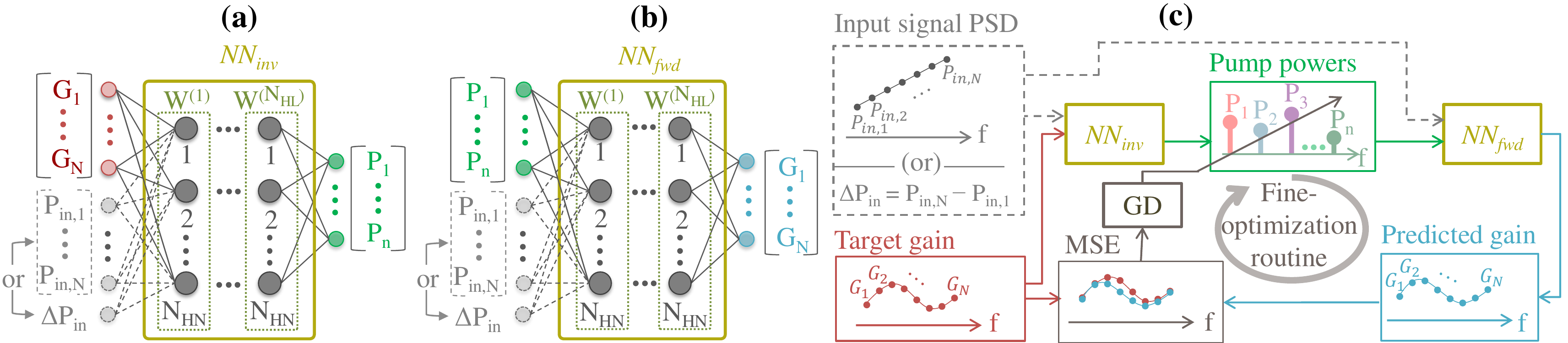}
\caption{Neural networks architectures for the (a) inverse ($NN_{inv}$) and (b) forward ($NN_{fwd}$) models with their respective inputs/outputs, (c) Machine learning framework and its modification (dashed lines) to support the input signal power spectral density (PSD) profile information in terms of $\mathbf{P_{in}}=[P_{in,1},\dots,P_\mathrm{in,N}]^T$ or just the scalar value $\Delta P_{in} = P_{in,N} - P_{in,1}$ for a linear PSD. GD: gradient descent, MSE: mean squared error.}
\label{fig:ML-FW}
\end{figure*}

In this work, we validate and modify the machine learning framework introduced by~\cite{Zibar:19,Zibar20} for the inverse Raman amplifier design. The modification consists in adding information about the input signal PSD profile into the NN models of the ML framework. Therefore, we characterize the robustness of the framework testing it experimentally for different Raman amplifier configurations and analyzing if the proposed modification will actually increase its accuracy. The new ML framework scheme is illustrated in Fig.~\ref{fig:ML-FW}(c) and works as following described.

The inverse design is performed by a neural network referred as $NN_{inv}$ and illustrated in Fig.~\ref{fig:ML-FW}(a). $NN_{inv}$ has previously learned the inverse mapping of the Raman amplifier. This inverse mapping is given by $\mathbf{P} = f^{-1}(\mathbf{G,P_{in}}$) or $\mathbf{P} = f^{-1}(\mathbf{G},\Delta P_{in})$, where $\mathbf{P}=[P_{1},\dots,P_\mathrm{n}]^T$ is a vector with $n$ pump powers, $\mathbf{G}=[G_{1},\dots,G_\mathrm{N}]^T$ and $\mathbf{P_{in}}=[P_{in,1},\dots,P_\mathrm{in,N}]^T$ are vectors with $N$ channelized points taken over the gain--profile curve and the input signal PSD profile, respectively, and $\Delta P_{in}$ is the difference between the powers of the highest and lowest frequencies of the channelized input signal PSD profile ($\Delta P_{in} = P_{in,N} - P_{in,1}$). Therefore, $\Delta P_{in}$ is a scalar value associated with the input signal PSD slope and it is useful when considering linear PSDs. After being properly trained, $NN_{inv}$ can instantly estimate the pump powers needed to achieve the target gain--profile (given a certain input signal PSD profile).

The $NN_{inv}$ output (pump power configuration) is tested using another NN, referred as $NN_{fwd}$ and illustrated in Fig.~\ref{fig:ML-FW}(b). $NN_{fwd}$ has learned the forward mapping of the Raman amplifier. This mapping is given by $\mathbf{G} = f(\mathbf{P,P_{in}})$ or $\mathbf{G} = f(\mathbf{P},\Delta P_{in})$. If the mean squared error (MSE) between target ($NN_{inv}$ input) and predicted ($NN_{fwd}$ output) gain--profiles is above a defined threshold, the pump powers can be fine--tuned by means of a gradient descent (GD) routine, as shown in Fig.~\ref{fig:ML-FW}(c). The convergence time of the GD fine--optimization routine is reduced since $NN_{inv}$ provides a good initialization point and $NN_{fwd}$ is a fast gain--profile predictor~\cite{RosaBrusin19}.

Regarding the input signal PSD profile awareness, henceforth, the $NN_{inv}$ and $NN_{fwd}$ models are referred as $\mathbf{P_{in}}$--aware when considering the additional input $\mathbf{P_{in}}$; $\Delta P_{in}$--aware when considering the additional input $\Delta P_{in}$; and $\mathbf{P_{in}}$--unaware when none of these inputs are considered.

\subsection{Experimental data--set generation}
\textbf{\label{sec:data-set_gen}}
To generate the experimental data--sets to train and test the NN models considered in this work, we measured $M$ gain--profiles on the experimental setup in Fig.~\ref{fig:setup}(a). These profiles correspond to different pump configurations drawn from uniform distributions. A pump configuration is associated to a single and linear input signal PSD profile. If this PSD profile is random, defined by $\Delta P_{in} \sim U[-5,+5]$~dB, the data--set is referred as DS--$\Delta P_{in}$--var. Instead, if the input signal PSD profile is the same for all pump configurations, the data--set (DS) is referred as DS--$\Delta P_{in}$--fixed. 

DS--$\Delta P_{in}$--var is considered only for the distributed Raman amplifier using SSMF. In this data--set, the total input power varies from 7 to 9 dBm, depending on the considered slope $\Delta P_{in}$. For the DS--$\Delta P_{in}$--fixed, the total input power remains constant, being around 9 dBm for SSMF, and ULLF and 3 dBm for HNLF, DCF and IDF.

To evaluate the ML framework performance over different fiber types, only the $\mathbf{P_{in}}$--unaware models are considered. These models are trained and tested over DS--$\Delta P_{in}$--fixed data--set without the input signal PSD information. In this case, the data--sets are given by $\mathcal{D}^{M\times (n+N)}=\{(P_1^i,...P_\mathrm{n}^i,G_1^i,...,G_\mathrm{N}^i)|i=1,...,M\}$, with $n$=~4 and $N$ =~40. On the other hand, the input profile impact over the ML framework considers all input signal PSD awareness models. $\mathbf{P_{in}}$--unaware models are trained as before. $\mathbf{P_{in}}$--aware and $\Delta P_{in}$--aware models are jointly trained over DS--$\Delta P_{in}$--fixed and DS--$\Delta P_{in}$--var data--sets. These data--sets are given by either $\mathcal{D}^{M\times (n+2N)}=\{(P_1^i,...P_\mathrm{n}^i,P_{in,1}^i,...P_{in,\mathrm{N}}^i,G_1^i,...,G_\mathrm{N}^i)|i=1,...,M\}$ or $\mathcal{D}^{M\times (n+1+N)}=\{(P_1^i,...P_\mathrm{n}^i,\Delta P_{in}^i,G_1^i,...,G_\mathrm{N}^i)|i=1,...,M\}$, depending on whether the entire $\mathbf{P_{in}}$ or just $\Delta P_{in}$ is considered as the input signal PSD information, respectively. All these models are separately tested over DS--$\Delta P_{in}$--fixed and DS--$\Delta P_{in}$--var.

Regarding the data--sets sizes, for each fiber type (except SSMF 100--km), a different DS--$\Delta P_{in}$--fixed data--set with $M$ = 3000 cases is generated. For SSMF 100--km, DS--$\Delta P_{in}$--fixed data--set has $M$ = 6000 and DS--$\Delta P_{in}$--var data--set has $M$ = 10000 cases. The experimental data--sets are split in two halves, referred as $\mathcal{D}_1$ and $\mathcal{D}_2$. $\mathcal{D}_1$ is used to train, test and validate the neural network models. $\mathcal{D}_2$ is used to experimentally evaluate the final overall performance of the ML framework. $\mathcal{D}_2$ is also used to retest the individual NN models in Sections~\ref{sec:inverse_model_performance} and ~\ref{sec:direct_model_performance}.

\subsection{Model selection and training}
\textbf{\label{sec:training_models}}
All $NN_{inv}$ models are trained using random projection (RP) (also known as extreme learning machine, ELM)~\cite{Huang2011}. This is a fast training algorithm that optimizes only the last layer weights by regularized least squares (regularization parameter $\lambda$). The hidden layers are randomly assigned according to a normal distribution (zero mean and a pre--defined standard deviation $\sigma_{NN}$). A hyperparameter optimization is performed to achieve good generalization properties. It considers a grid search procedure applying 10-fold cross validation where 90\% of $\mathcal{D}_1$ is reserved for training and 10\% for validation. The hyperparameters optimized in this work are the activation function ($f_{act}$), the number of hidden nodes ($N_{HN}$), the number of hidden layers ($N_{HL}$), $\sigma_{NN}$ and $\lambda$. The impact of the randomly initialized weights is reduced by training $N_{NN}$ parallel and independent $NN_{inv}$. Therefore, the pump configuration prediction will be the average of the $N_{NN}$ distinct $NN_{inv}$ outputs~\cite{Zibar20}. $NN_{fwd}$ models, on the other hand, are trained using the Levenberg-Marquadt (LM) method. In this case, $\mathcal{D}_1$ was split in three parts with 70\%, 15\% and 15\% of the data used for training, validation and test, respectively. All the hyperparameters for $NN_{inv}$ and $NN_{fwd}$ are summarized in Table~\ref{tab:NNmodels}.

\begin{table*}[!h]
\centering
\caption{\bf Neural network parameters for the different models}
\begin{tabular}{lllllllllll}
\hline
Models & Input signal PSD awareness & Optical fiber & Training algorithm & $f_{act}$    & $N_{HL}$  & $N_{HN}$  & $N_{NN}$  & $\sigma_{NN}$ & $\lambda$ \\
\hline
$NN_{inv}$ & $\mathbf{P_{in}}$--unaware             & all fibers\textsuperscript{1} & RP & tanh     & 2 & 600   & 10    & 5e-2      & 1e6       \\
$NN_{inv}$ & $\mathbf{P_{in}}$--unaware             & SSMF 100-km                   & RP & logsig   & 1 & 1300  & 20    & 5e-2      & 1e-8      \\
$NN_{inv}$ & $\mathbf{P_{in}}$--aware               & SSMF 100-km                   & RP & sine     & 1 & 1000  & 20    & 7e-3      & 1e-8      \\
$NN_{inv}$ & $\Delta P_{in}$--aware                 & SSMF 100-km                   & RP & sine     & 1 & 1000  & 20    & 2.5e-2    & 1e-8      \\
$NN_{fwd}$ & (all $\mathbf{P_{in}}$ awareness)      & all fibers                    & LM & tanh     & 2 & 10    & 1     & See\textsuperscript{2} & See\textsuperscript{3} \\
\hline
\end{tabular}
\vspace{1ex}
{\\ \raggedright 1 Except SSMF 100-km; 2 Nguyen-Widrow initialization algorithm~\cite{initnw}; 3 Dynamically modified during training according to~\cite{Hagan94}. \par}
  \label{tab:NNmodels}
\end{table*}

\subsection{Inverse models performance}
\label{sec:inverse_model_performance}
The inverse models accuracy in predicting the pump powers are shown in terms of maximum error between true and predicted power values for each pump laser over the test data--sets $\mathcal{D}_2$. The results in Table~\ref{tab:NNinv_perform_fibers} are for different fiber types considering the $\mathbf{P_{in}}$--unaware models tested over DS--$\Delta P_{in}$--fixed data--set only. Table~\ref{tab:NNinv_perform_profiles} presents the results for SSMF 100--km considering all input signal PSD awareness models: $\mathbf{P_{in}}$--unaware, $\mathbf{P_{in}}$--aware and $\Delta P_{in}$--aware, where each model is separately tested over DS--$\Delta P_{in}$--fixed and DS--$\Delta P_{in}$--var data--sets.

\begin{table}[h]
\centering
\caption{\bf \boldmath $NN_{inv}$ $\mathbf{P_{in}}$--unaware performance: maximum error between target and predicted pump powers for different fiber types over their respective DS--$\Delta P_{in}$--fixed data--set}
\begin{tabular}{ccccccc} 
\hline
Fiber           		& HNLF   & DCF 	& IDF 		& SSMF & ULLF   \\
Length (km)     		& 7.5    & 4.8 	& 15 		& 50   & 50     \\
\hline
$P_1$ (mW) & 28.6 & 17.2 & 31.2 & 19.3 & 50.3 \\
$P_2$ (mW) & 45.8 & 19.9 & 47.7 & 38.4 & 47.7 \\
$P_3$ (mW) & 96.5 & 19.8 & 60.7 & 22.1 & 39.4 \\
$P_4$ (mW) & 98.5 & 29.4 & 93.3 & 25.5 & 9.6 \\
\hline
\end{tabular}
  \label{tab:NNinv_perform_fibers}
\end{table}

In Table~\ref{tab:NNinv_perform_fibers}, errors higher than 50~mW (but below 100 mW) are observed just for HNLF and IDF ($P_3$ and $P_4$). These errors can lead to gain variations from 1 to 2 dB. The other fiber types have accurate prediction performance with maximum errors below 50~mW. These pump errors correspond to low gain variations of up to 1~dB. The worse performance for HNLF and IDF might be related to their higher measured gains when compared to the other fiber types as shown in Fig.~\ref{fig:setup}(d-i).

\begin{table}[h]
\centering
\caption{\bf \boldmath $NN_{inv}$ SSMF 100--km performance: maximum error between target and predicted pump powers over DS--$\Delta P_{in}$--fixed and DS--$\Delta P_{in}$--var data--sets for different input signal PSD awareness}
\begin{tabular}{ccccccc} 
\hline
    & \multicolumn{2}{c}{$\mathbf{P_{in}}$--unaware} & \multicolumn{2}{c}{$\mathbf{P_{in}}$--aware} & \multicolumn{2}{c}{$\Delta P_{in}$--aware} \\
DS--$\Delta P_{in}$-- & fixed   & var  & fixed  & var & fixed & var \\
\hline
$P_1$ (mW) & 15.1 & 41.5 & 11.8 & 15.4 & 13.5 & 26.5 \\
$P_2$ (mW) & 22.6 & 90.2 & 17.6 & 27.9 & 21.3 & 39.0 \\
$P_3$ (mW) & 33.6 & 55.9 & 27.1 & 20.8 & 22.0 & 23.7 \\
$P_4$ (mW) & 41.1 & 60.8 & 38.2 & 27.9 & 31.7 & 41.6 \\
\hline
\end{tabular}
  \label{tab:NNinv_perform_profiles}
\end{table}

For the SSMF 100--km, the results in Table~\ref{tab:NNinv_perform_profiles} show that by training without the information of the signal input PSD profile, $\mathbf{P_{in}}$--unaware $NN_{inv}$ model can only have accuracy when tested over the DS--$\Delta P_{in}$--fixed data--set (errors below 50~mW). Its performance over DS--$\Delta P_{in}$--var data--set can reach 90~mW of error for $P_2$. These results are not intuitive since it is expected that input signal PSD variations should not affect the gain--profile since signal--signal and pump--pump SRS interactions are not so strong in the C--band. However, the inverse model proved to be very sensitive even for small SRS effects. $\mathbf{P_{in}}$--aware and $\Delta P_{in}$--aware $NN_{inv}$ models, on the other hand, are able to maintain the same accuracy for both DS--$\Delta P_{in}$--fixed and DS--$\Delta P_{in}$--var data--sets since they consider the information of the signal input PSD profile.

\begin{figure}[h]
  \centering
  \includegraphics[width=0.49\textwidth]{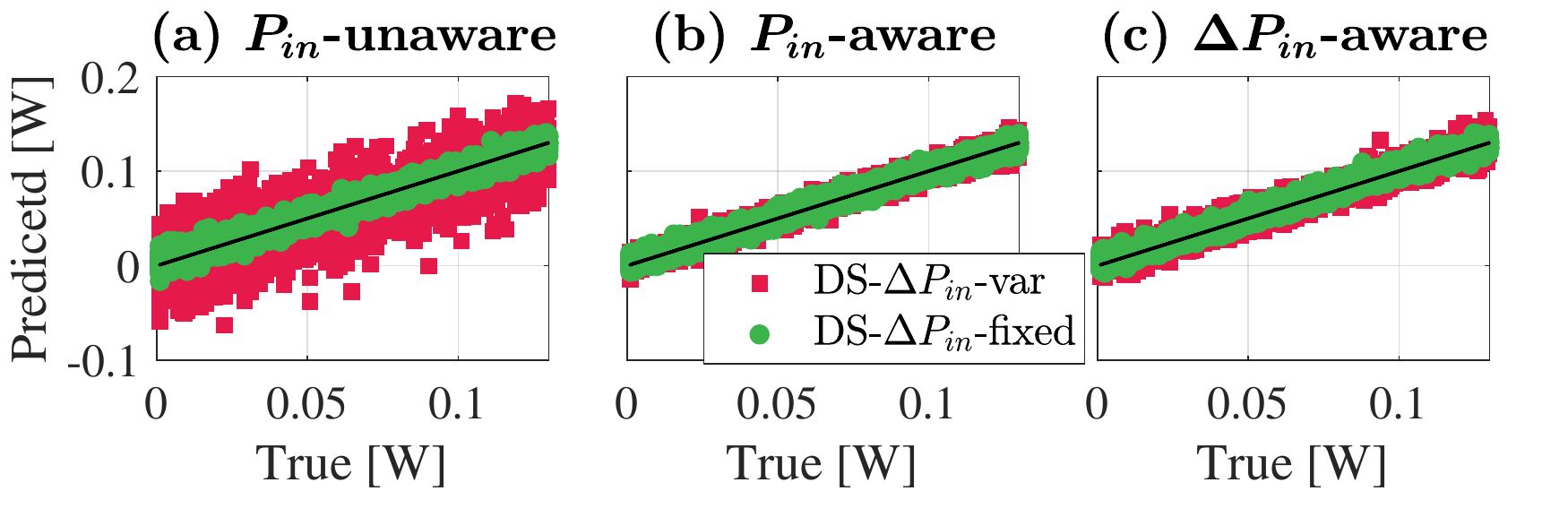}
\caption{Inverse model performance in terms of predicted and target pump power (pump 2) over DS--$\Delta P_{in}$--fixed and DS--$\Delta P_{in}$--var test data--sets.}
\label{fig:NNinv_perform_profiles}
\end{figure}

This is also shown in Fig.~\ref{fig:NNinv_perform_profiles}, which illustrates the inverse model performance when predicting pump laser $P_2$ (the pump with the lowest accuracy for the $\mathbf{P_{in}}$--unaware over DS--$\Delta P_{in}$--var). The similar performance for $\mathbf{P_{in}}$--aware and $\Delta P_{in}$--aware is promising since it indicates that for this specific analysis of linear input signal PSD--profiles, the $\Delta P_{in}$ information is sufficient to provide highly accurate pump predictions. However, more arbitrary input signal PSD shapes (nonlinear) would still need the entire vector $\mathbf{P_{in}}$ since a single scalar value is not able to define a more general input signal PSD shape.

\subsection{Forward models performance}
\label{sec:direct_model_performance}
\begin{figure}[h]
  \centering
  \includegraphics[width=0.49\textwidth]{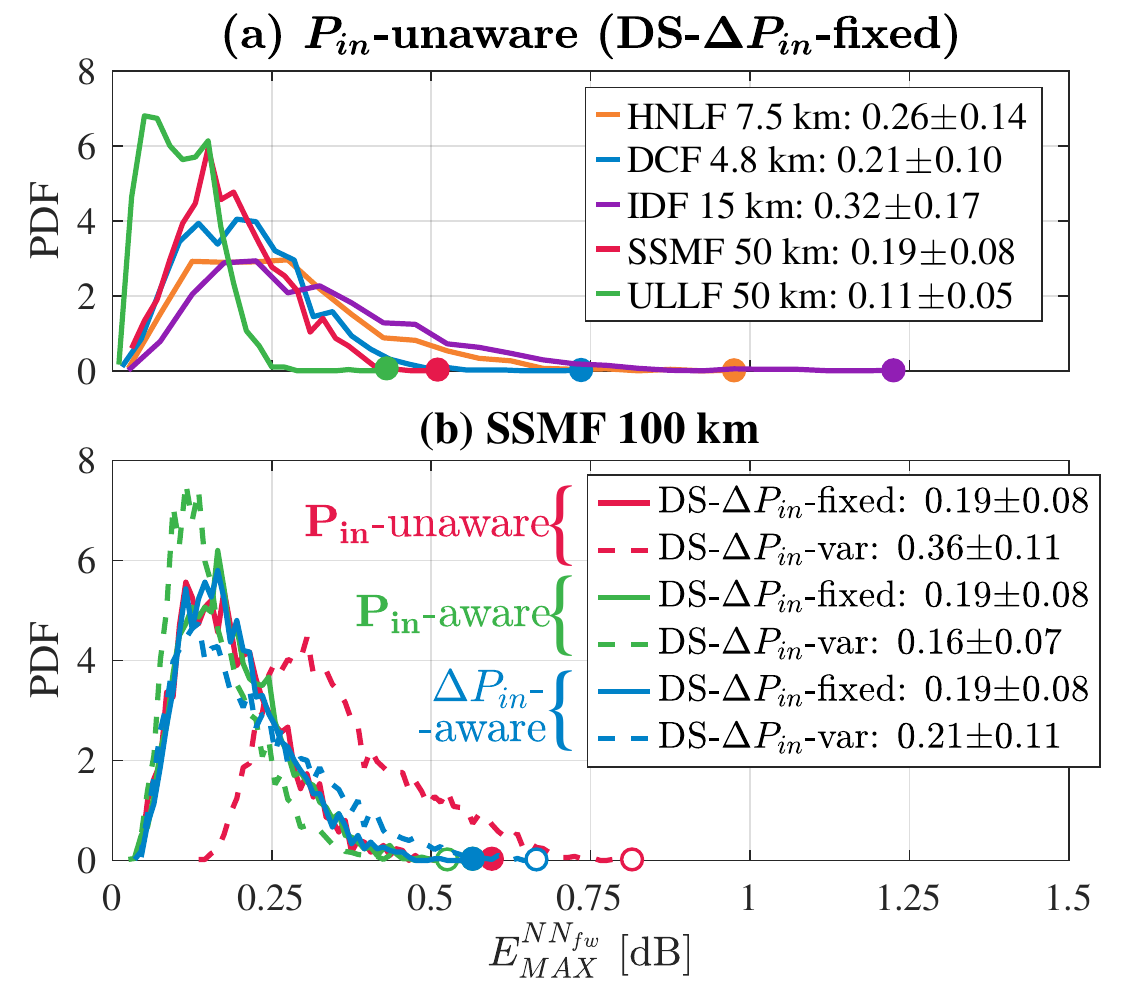}
\caption{Forward model performance in terms of probability density function (PDF) for the maximum error $E^{NN_{fw}}_{MAX}$ when predicting the gain-profile for (a) different optical fibers $\mathbf{P_{in}}$--unaware models over DS--$\Delta P_{in}$--fixed test data--set and (b) SSMF 100-km with different input signal PSD awareness over DS--$\Delta P_{in}$--fixed and DS--$\Delta P_{in}$--var test data--sets. Dots report the maximum $E^{NN_{fw}}_{MAX}$ values.}
\label{fig:NNdir_perform}
\end{figure}

To evaluate the prediction accuracy of the forward models, the maximum absolute error $E^{NN_{fw}}_{MAX}$ between target and predicted gain--profiles along the frequency is calculated for the cases on the test data--sets $\mathcal{D}_2$. A statistical analysis of $E^{NN_{fw}}_{MAX}$ in terms of probability density function (PDF) is performed for each evaluated case. The PDF curves are shown in Fig.~\ref{fig:NNdir_perform}(a) for the models of different fiber types and in Fig.~\ref{fig:NNdir_perform}(b) for the SSMF 100-km considering different input signal PSD awareness models. Mean ($\mu$) and standard deviation ($\sigma$) values are also reported as $\mu \pm \sigma$.

As observed for the $NN_{inv}$, the forward models for HNLF and IDF present the worst prediction performance when compared to the other fiber types. However, all forward models in Fig.~\ref{fig:NNdir_perform}(a) are highly accurate, with mean ($\mu$) of $E^{NN_{fw}}_{MAX}$ below 0.32 dB and maximum values lower than 1.25~dB.

The results for the SSMF 100-km shown in Fig.~\ref{fig:NNdir_perform}(b) are also similar to the inverse model performance. Again, the $\mathbf{P_{in}}$--unaware model present a worse prediction performance over the DS--$\Delta P_{in}$--var when compared to the DS--$\Delta P_{in}$--fixed data--set. However, this degradation is negligible, with a mean $E^{NN_{fw}}_{MAX}$ decrease of only 0.17~dB and a maximum $E^{NN_{fw}}_{MAX}$ going from 0.6 to 0.8~dB. Therefore, different for the inverse model, the forward model seems to be more robust the small SRS effects between signals on the C--band. $\mathbf{P_{in}}$--aware and $\Delta P_{in}$--aware models, again, are able to provide highly gain--profile prediction accuracy for both DS--$\Delta P_{in}$--fixed and DS--$\Delta P_{in}$--var data--sets.

\section{Results and discussions}
\label{sec:results}
To validate the ML framework on designing/achieving a target Raman gain profile, a new set of measurements is taken. The procedure consists in feeding the target gain--profile into the ML framework in Fig.~\ref{fig:ML-FW}(c), configuring the pump powers provided by it on the experimental setup in Fig.~\ref{fig:setup}(a), and measuring the gain. Target and measured gain--profiles are compared by means of maximum absolute error $E_{MAX}$ over frequency.

Two sets of target gain--profiles are investigated: arbitrary and flat/tilted. The gain profiles on the test data--set $\mathcal{D}_2$ are used as the target arbitrary gains. These gains are achievable profiles for the considered experimental setup. By using them as target gains it is possible to evaluate the ML framework for both: generalization (since these gains were not used on the NN models training stage) and accuracy in terms of how close it can provide feasible gain--profiles. Flat and tilted gain--profiles realization, on the other hand, are not guaranteed on the available experimental setup. This way, the ML framework is tested on its handling of a broader generalization problem, where the target may only be approximated.

\begin{figure*}[t]
  \centering
  \includegraphics[width=1.00\textwidth]{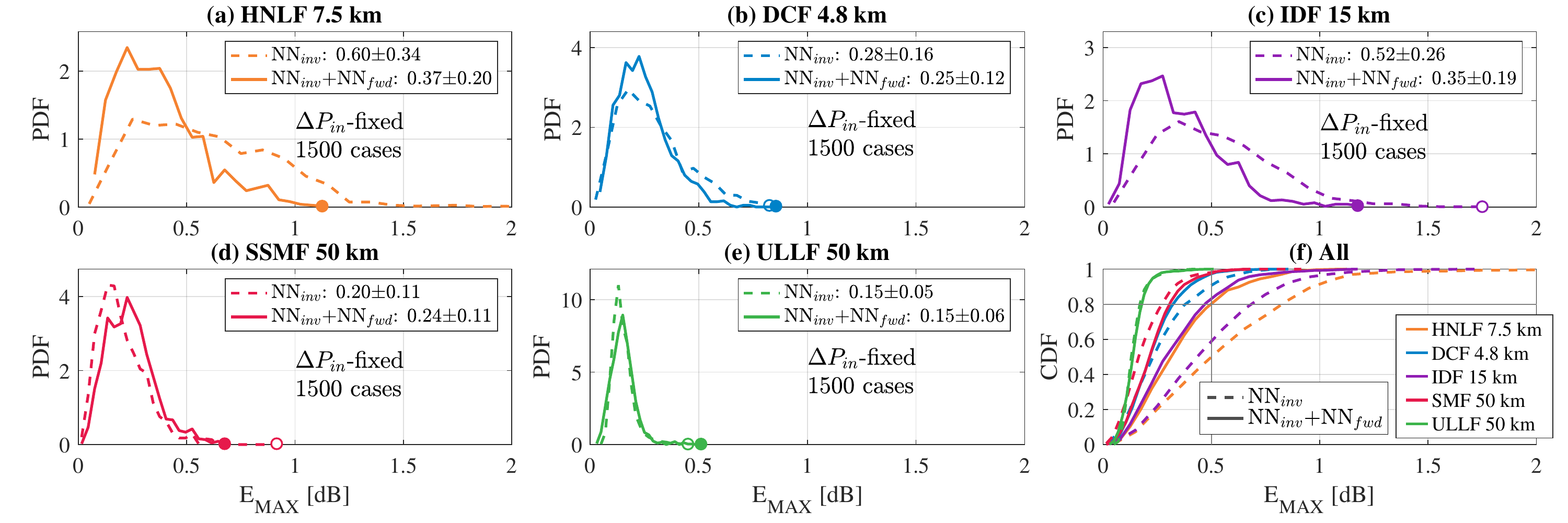}
\caption{Probability density function (PDF) and cumulative distribution function (CDF) of the maximum absolute error ($E_{MAX}$) between target and measured arbitrary gain--profiles (DS--$\Delta P_{in}$--fixed test data--set) for the $\mathbf{P_{in}}$--unaware models.}
\label{fig:pdf_result_allFibers}
\end{figure*}

Fig.~\ref{fig:pdf_result_allFibers} shows $E_{MAX}$ PDFs and cumulative distribution functions (CDF) over the arbitrary gains and for each evaluated fiber type. Two curves are shown: $NN_{inv}$ (dashed lines), when only applying $NN_{inv}$, and $NN_{inv}+NN_{fwd}$ (solid lines), when employing the GD fine--optimization routine using the $NN_{fwd}$ estimation. Mean ($\mu$) and standard deviation ($\sigma$) values are also reported. Only HNLF (Fig.~\ref{fig:pdf_result_allFibers}(a)) has cases with $E_{MAX}$ higher than 2~dB. But since these cases are just a few (0.47\% of the cases), the x--axis limit is set to 2~dB for all evaluated fiber types.

The results in Fig.~\ref{fig:pdf_result_allFibers} show that for highly nonlinear fibers such as HNLF (Fig.~\ref{fig:pdf_result_allFibers}(a)) and IDF (Fig.~\ref{fig:pdf_result_allFibers}(c)), the design applying $NN_{inv}$ can only provide a moderate accuracy in experimentally realizing the target gain--profiles. For these fibers, just 50\% and 59\% of the cases have $E_{MAX}$ $<$ 0.5~dB, according to their CDF curves in Fig.~\ref{fig:pdf_result_allFibers}(f). These results are expected since these fibers present the worse inverse model performance as shown in Section~\ref{sec:inverse_model_performance}.

Although DCF (Fig.~\ref{fig:pdf_result_allFibers}(b)) is also a highly nonlinear fiber, it provides overall lower gain (see Fig. \ref{fig:setup}(e)) and the $NN_{inv}$ presents a significantly better accuracy with 91\% of the cases with $E_{MAX}$ below 0.5~dB as shown in Fig.~\ref{fig:pdf_result_allFibers}(d). High design accuracy is also observed for SSMF (Fig.~\ref{fig:pdf_result_allFibers}(d)) and ULLF (Fig.~\ref{fig:pdf_result_allFibers}(e)) when applying $NN_{inv}$. For these two fibers, the CDF curves have $E_{MAX}$ below 0.5~dB for 98\% (SSMF 50~km) and 100\% (ULLF) of the cases. Again, all these fibers have a better inverse model accuracy (Section~\ref{sec:inverse_model_performance}), which explains why their inverse designs have a better performance when comparing to HNLF and IDF.

By considering the GD fine--optimization routine ($NN_{inv}+NN_{fwd}$ curves), the performance for HNLF and IDF significantly increase. Their CDF curves in Fig.~\ref{fig:pdf_result_allFibers}(f) shows that now 80\% (HNLF) and 83\% (IDF) of the cases have $E_{MAX}<$ 0.5~dB. Recall that their $NN_{fwd}$ (Section \ref{sec:direct_model_performance}) have high accuracy and, therefore, are able to optimize the pump powers provided by $NN_{inv}$. For the other fiber types (DCF, SSMF and ULLF), on the other hand, the GD fine--optimization routine does not introduce important changes. This is because, for these cases, the $NN_{inv}$ outcome is already on a local minimum and the GD just add random deviations around it due to $NN_{fwd}$ prediction errors.

\begin{figure*}[t]
  \centering
  \includegraphics[width=1.00\textwidth]{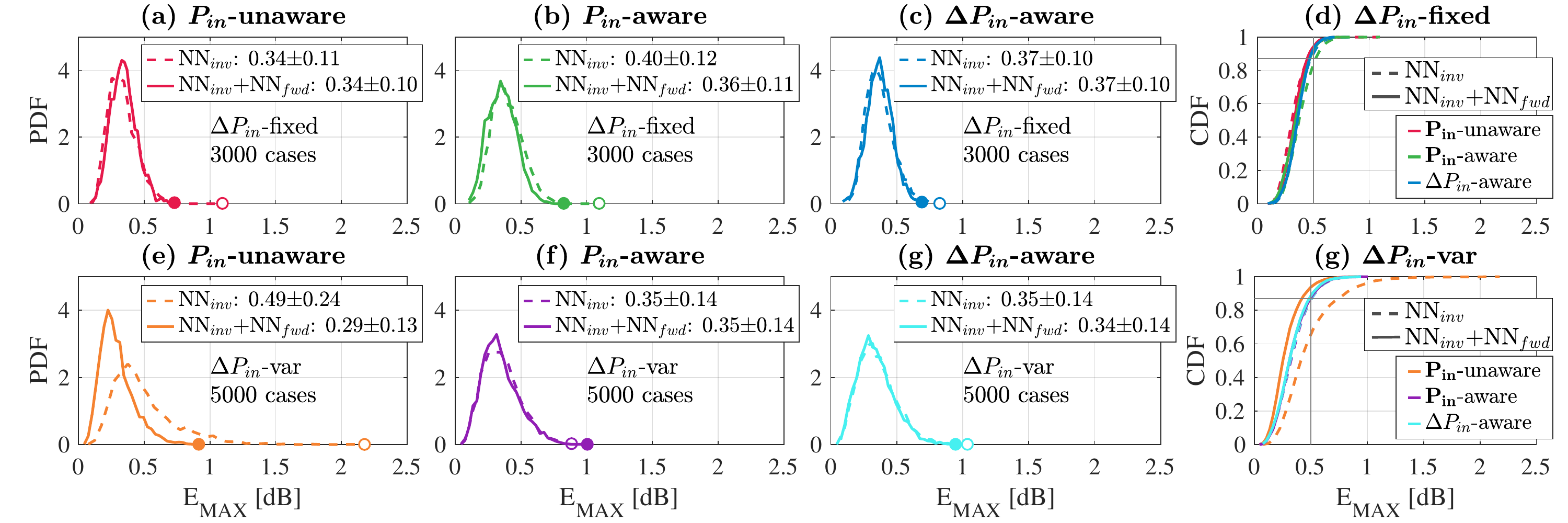}
\caption{SSMF 100-km: probability density function (PDF) and cumulative distribution function (CDF)  of the maximum absolute error ($E_{MAX}$) between target and measured gain--profiles over the test DS--$\Delta P_{in}$--fixed and DS--tilt  for the $\mathbf{P_{in}}$--unaware, $\mathbf{P_{in}}$--aware and $\Delta P_{in}$--aware models.}
\label{fig:pdf_result_all_power_profiles}
\end{figure*}

The results for SSMF 100-km are separately shown in Fig.~\ref{fig:pdf_result_all_power_profiles}. Fig.~\ref{fig:pdf_result_all_power_profiles}(a-d) consider DS--$\Delta P_{in}$--fixed data--sets and Fig.~\ref{fig:pdf_result_all_power_profiles}(e-f) consider DS--$\Delta P_{in}$--var data--sets. When only applying $NN_{inv}$ (dashed lines), highly accurate performance are observed for all models and data--sets, except for $\mathbf{P_{in}}$--unaware model tested over DS--$\Delta P_{in}$--var data--set (Fig.~\ref{fig:pdf_result_all_power_profiles}(e)). For this case, $E_{MAX}$ values are up to 2.2~dB. CDF curves in Fig.~\ref{fig:pdf_result_all_power_profiles}(g) shows that around 70\% of the cases have errors below 0.5~dB for the $\mathbf{P_{in}}$--unaware model, while for all the other models it occurs for 80\% of the cases. This is expected since the pump prediction is more degraded according to the results presented in Section~\ref{sec:inverse_model_performance} for $\mathbf{P_{in}}$--unaware inverse model $NN_{inv}$.

When applying the GD--based fine--optimization (solid lines), a significant improvement is observed only for the $\mathbf{P_{in}}$--unaware model over the DS--$\Delta P_{in}$--var data--set, as shown in Fig.~\ref{fig:pdf_result_all_power_profiles}(e). Instead, for all the other models and data--sets, GD fine--optimization does not provide significant improvements. Again, for these cases, the pump configuration provided by $NN_{inv}$ might be already on a local minimum. Thus, the GD fine--optimization only randomly disturb it. Moreover, the best performance of $\mathbf{P_{in}}$--unaware against $\mathbf{P_{in}}$--aware and $\Delta P_{in}$--aware models when applying the GD fine--optimization might be related to its lower input dimension. Recall that larger neural networks have more parameters and get easily trapped in a local minimum~\cite{li2017visualizing}.

Overall, when designing arbitrary gain---profiles, no matter the fiber type or the input signal PSD condition when training the NN models, the ML framework is able to maintain more than 80\% of the cases with errors below 0.5~dB by just considering the $\mathbf{P_{in}}$--unaware models. Moreover, for DCF, SSMF and ULLL this high accuracy is achieved by just applying $NN_{inv}$, leading to a non--iterative gain--profile adjustments relying on matrix multiplications. This is also true for the $\mathbf{P_{in}}$--aware and $\Delta P_{in}$--aware models. With the last being a very interesting and low complex approach that increases the NN dimension by just one additional input when considering linear PSD profiles.

Next, the ability of the ML framework to provide accurate flat and tilted gain--profiles for the different fiber types is investigated. This analysis is taken using $\mathbf{P_{in}}$--unaware models trained over constant input signal PSD (DS--$\Delta P_{in}$--fixed data--set). Flat and tilted gains--profiles ranging from 1 to 6~dB are evaluated in steps of 1~dB. For the tilted profiles, negative and positive slopes of 1~dB over the C--band are considered.

\begin{figure*}[t]
  \centering
  \includegraphics[width=1.00\textwidth]{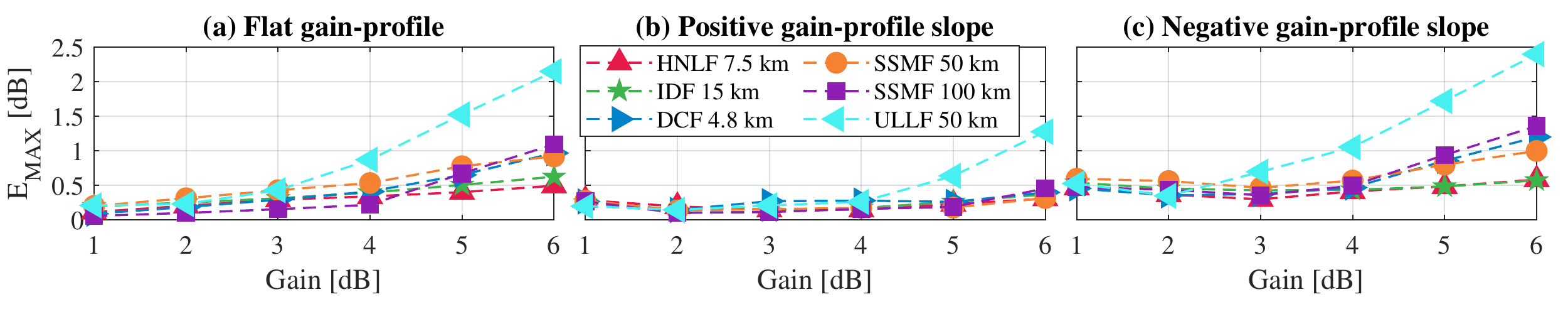}
\caption{Errors for (a) flat, (b) positive-- and (c) negative--slope tilted gain--profiles after fine--optimization for the evaluated fiber types considering the $\mathbf{P_{in}}$--unaware models.}
\label{fig:flat_tilted_errors_result_allFibers}
\end{figure*}

Fig.~\ref{fig:flat_tilted_errors_result_allFibers} summarizes the results for all evaluated fiber types also in terms of $E_{MAX}$. These results are obtained after applying the GD--based fine--optimization ($NN_{inv}+NN_{fwd}$), since it presents a better performance in achieving flat/tilted profiles when compared to just applying the $NN_{inv}$. 

For the flat gain--profiles in Fig.~\ref{fig:flat_tilted_errors_result_allFibers}(a), $E_{MAX}$ values are kept below 1~dB for all fibers, except for the ULLF. Notice that, due to its lower Raman gain coefficient, ULLF can provide gains of up to 4~dB on the low frequency region according to the measured gains over the data--set (see Fig.~\ref{fig:setup}(i)). Therefore, it is not possible to have flat gains higher than 4~dB given our pump power limits constraint. Recall the ULLF 50-km is the fiber with the best performance in achieving arbitrary gains. This is because the arbitrary gains are taken from the test data--set, being achievable considering the available pump powers.

Tilted gain--profiles with a positive slope (Fig.~\ref{fig:flat_tilted_errors_result_allFibers}(b)) present a better performance when compared to the flat slope. An almost equally spaced pump configuration (in frequency), as the one used in this work, tends to provide a negative gain--profile slope due to the power transferred from high to low frequencies channels~\cite{Rottwitt99}. However, for the narrow C--band considered in this work, SRS between channels (and also between pumps) are not so strong to provide a negative gain slope due to the wide Raman peak ($\sim$ 13 THz). Thus, the final gain profile resulted from these pumps are more influenced by the superposition of the Raman gain curves of each individual pump lasers, which leads to a positive, rather than a negative gain--profile slope (see Fig.~\ref{fig:setup}(c-i)). This also explains the worse performance observed for the negative gain--profile slope (Fig.~\ref{fig:flat_tilted_errors_result_allFibers}(c)) when compared to the positive ones (Fig.~\ref{fig:flat_tilted_errors_result_allFibers}(b)). Therefore, positive gain--profile slopes have a better performance because they benefit from the pump frequency distribution. Positive slopes have $E_{MAX}$ below 0.5~dB for all fibers, while for negative slopes $E_{MAX}$ can reach 1.5~dB. This discussion excludes again the ULLF. The ULLF-based RA operating under high gain suffers from significantly higher error values, due to the higher pump power that would be required compared to the power available in our experimental setup (referred to as ``power limitation'').

\begin{figure*}[t]
  \centering
  \includegraphics[width=1.0\textwidth]{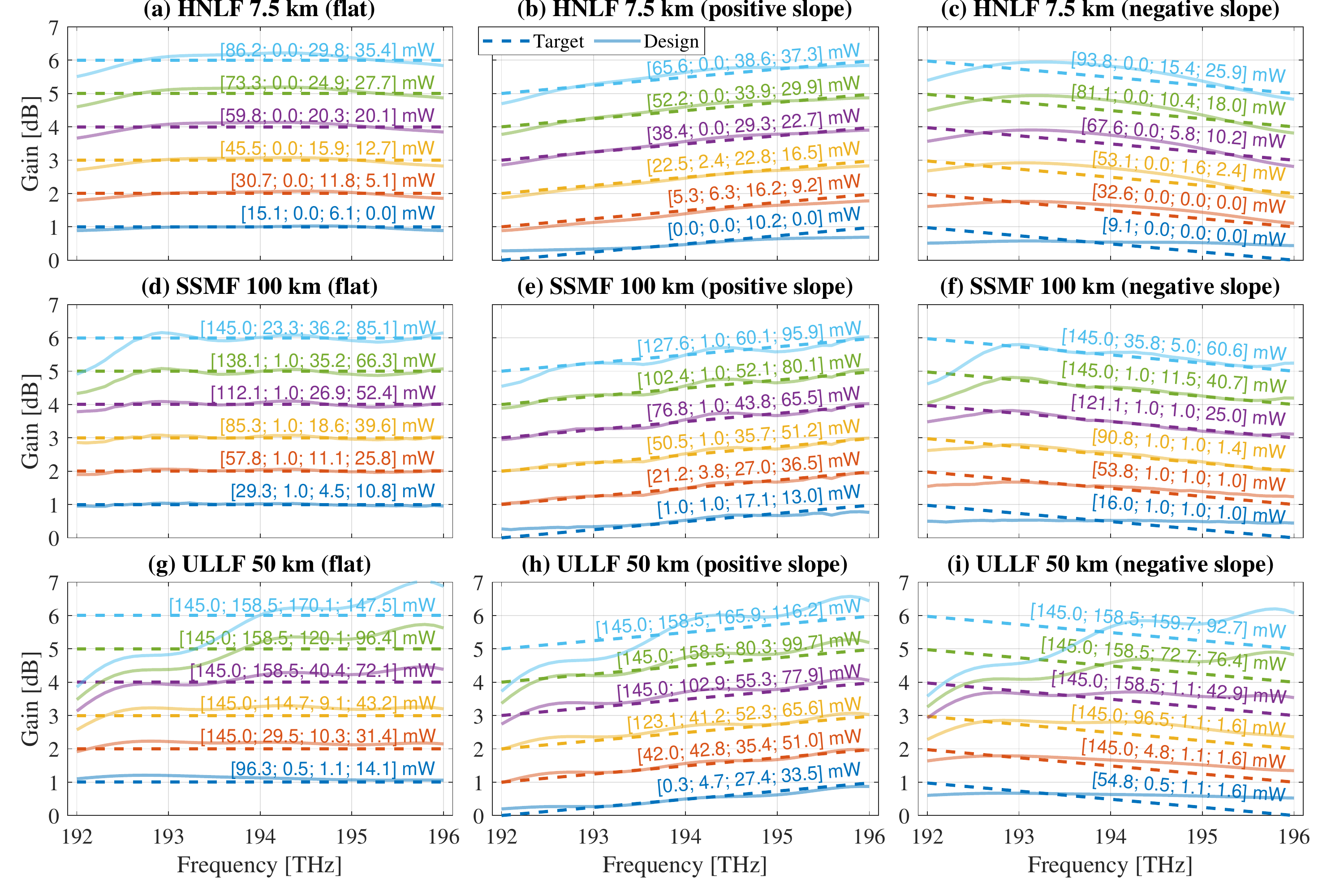}
\caption{Target and measured gain--profiles after fine--optimization showing the pump power values [$P_1$, $P_2$, $P_3$, $P_4$] for HNLF with (a) flat, (b) positive-- and (c) negative--slope tilted gain--profiles; SSMF 100--km with (d) flat, (e) positive and (f) negative slopes tilted gain--profiles; and ULLF with (g) flat, (h) positive and (i) negative slopes tilted gain--profiles.}
\label{fig:Experimental_HNLFandULLF}
\end{figure*}

Fig.~\ref{fig:Experimental_HNLFandULLF} shows target and measured gain--profile curves for three fiber types. Fig.~\ref{fig:Experimental_HNLFandULLF}(a-c) is for HNLF with the highest Raman gain coefficient and the best performance in achieving flat/tilted gain--profiles; Fig.~\ref{fig:Experimental_HNLFandULLF}(d-f) is for SSMF 100-km, which is the most commonly used for distributed Raman amplifiers; and Fig.~\ref{fig:Experimental_HNLFandULLF}(g-i) is for ULLF, with the lowest Raman gain coefficient and the worst performance in achieving flat/tilted gain--profiles. The pump power configurations provided by the ML framework are also reported. 

The HNLF has the best performance, with just three pumps ($P_1$, $P_3$ and $P_4$) being necessary to achieve a highly accurate flat and tilted gain--profiles. The errors are more related to our available Raman amplifier, in terms of number of pumps and their fixed wavelengths than to pump power limitations, since no pump power has reached its limit (Table \ref{tab:pumps}).

For SSMF 100-km, again the same three pumps are needed to achieve most target gain--profiles. For 6~dB of target flat gain--profile (Fig.~\ref{fig:Experimental_HNLFandULLF}(d)), $P_1$ achieves its maximum power. At this point, to compensate for $P_1$ power limitations, the ML framework starts to turn $P_2$ on (although with a very low power that does not influence the final gain). The same occurs for the two highest negative gain--profile slopes (Fig.~\ref{fig:Experimental_HNLFandULLF}(f)). For these cases, a higher deviation from the target is observed for low frequency signals, the region where $P_1$ should provide gain. Thus, in these cases, the errors are associated to pump power constraints, which affect more the high gain levels.

More intense consequences of the pump power limitation are observed for the low Raman gain coefficient ULLF, showing high limitations in manipulating the gain shape given our experimentally limited pump powers. For this fiber, all pumps are needed to achieve gain--profiles higher than 2~dB. For flat and negative gain--profile slopes, $P_1$ already reaches its limits for 2~dB of target gain, and $P_2$ for 4~dB. For positive slopes, these limits are achieved for slightly higher gains. But for all cases, the need for higher pumps powers limits the performance in providing flat/tilted gain--profile for higher gain levels.

In Fig.~\ref{fig:flat_tilted_errors_result_allFibers}, the positive slope gains have a better performance than the negative ones: this is mainly because the pump frequency distribution and the C--band operation (low inter--signal SRS) tends to provide a positive gain profile rather than negative. This can be verified by the gains mismatch in Fig.~\ref{fig:Experimental_HNLFandULLF}(c,f,i) for the gain cases where the pump limits are not reached (all cases in Fig.~\ref{fig:Experimental_HNLFandULLF}(c), first 4 cases in Fig.~\ref{fig:Experimental_HNLFandULLF}(f) and first case in Fig.~\ref{fig:Experimental_HNLFandULLF}(i)). For these cases, the gain shape is controlled mostly by adjusting $P_1$ only. The other pumps are turned off or with a low power because they will increase the gains of the high frequency channels as well. $P_1$ cannot be increased because it will also increase the gain on the middle of the spectrum (from 193 to 195~THz).

Therefore, when operating under pump power limits, the ML--FM is able to generalize and experimentally provide gain--profiles really close to completely new targets. The performance in this region is limited just by total number of pump lasers and their fixed wavelengths, presenting errors of up to 0.5~dB.

\section{Conclusion}
\label{sec:conclusion}
In this work, the machine learning framework for the Raman amplifier inverse design was extensively tested. The ability to provide arbitrary, flat and tilted gain--profiles was evaluated for different experimental realizations of the Raman amplifier, considering different optical fiber types, and both discrete and distributed amplifiers. For this analysis, results showed that more than 80\% of the 1500 target arbitrary gain spectra were achieved with a maximum error below 0.5~dB. For the flat/tilted gain profiles, maximum errors are up to 0.5~dB for all cases allowed by the available pump power within our experimental setup. 

Moreover, we also evaluate the machine learning robustness over different input power spectral distributions. Results show that the worse performance of the machine learning framework in this new scenario can be improved in two ways. One possibility is by considering the already proposed fine--design routine based on a gradient--descent optimization on a model not requiring the signal profile information. Another option is to incorporate this information on the training phase, creating input signal aware models as the ones developed in this work, and avoiding the iterative optimizations.

Overall, this experimental analysis has proven that the machine learning framework is a versatile tool able to experimentally provide highly accurate designs for different and practical scenarios covering a wide range of target gain--profiles.

\section*{Acknowledgment}
We thank OFS Fitel Denmark for providing the ultra--low loss fiber (SCUBA) used in this work. This project has received funding from the European Research Council through the ERC-CoG FRECOM project (grant agreement no. 771878), the European Union's Horizon 2020 research and innovation programme under the Marie Sk\l{}odowska-Curie grant agreement No 754462 and the Villum Foundations (VYI OPTIC-AI grant no. 29344).

\ifCLASSOPTIONcaptionsoff
  \newpage
\fi

\bibliographystyle{IEEEtran}
\bibliography{IEEEabrv,bibliography}

\end{document}